\documentclass[12pt,preprint]{revtex4}
\usepackage{graphicx}

\begin{document}

\preprint{Preprint option}
\title{The Effect of Dissipation on the Torque and Force Experienced by
Nanoparticles in an AC Field}
\author{F. Claro}
\email{fclaro@uc.cl}
\affiliation{Pontificia Universidad Cat\'olica de Chile, Casilla 306, Santiago, Chile }
\author{R. Fuchs}
\email{fuchs@iastate.edu}
\affiliation{Ames Laboratory and Iowa State University, Ames, Iowa 50011, USA}
\author{P. Robles}
\email{probles@ucv.cl}
\affiliation{Escuela de Ingenier\'{\i}a El\'ectrica, Pontificia Universidad Cat\'olica de
Valpara\'{\i}so, Casilla 4059, Valpara\'{\i}so, Chile}
\author{R. Rojas}
\email{roberto.rojas@usm.cl}
\affiliation{Universidad T\'ecnica Federico Santa Mar\'{\i}a, Casilla 110-V, Valpara\'{\i}%
so, Chile}

\begin{abstract}

We discuss the force and torque acting on spherical particles in an ensemble in the presence of
a uniform AC electric field.  We show that for a torque causing particle rotation to appear
the particle must be absorptive. Our proof includes all
electromagnetic excitations, which in the case of two or more particles
gives rise to one or more resonances in the spectrum of force and torque depending on
interparticle distance. Several peaks are found in the force and torque between two spheres
at small interparticle distances, which coalesce to just one as the separation grows
beyond three particle radii. We also show that in the presence of dissipation
the force on each particle is non conservative and may not be
derived from the classical interaction potential energy as has been done in the past.
\end{abstract}

\maketitle

\section{introduction}

\label{sec:intro}

The electromagnetic excitations and ensuing dynamics of nanoparticles, molecules and atoms in the presence of an
electric field has been widely studied both theoretically and
experimentally \cite{HXu,AHal,VPDr,YZh, McA}. The particles may initially be
unpolarized, but due to the external field and their mutual interaction they
may acquire induced dipole and higher electric moments.
As a consequence electric forces and torques are produced, resulting in
particle motion and the formation of equilibrium configurations.
An important case is optical trapping and binding, which, if many
particles are involved may lead to self-assembly of
ordered structures \cite{TCi}.
Structures are also formed in electro-rheological fluids, where a static
or slowly varying field induces the formation of linear arrays and columns
in a medium containing polarizable spheres in suspension \cite{Ta}.
Examples where an understanding of forces and torques is also crucial are
the dielectrophoresis and electrorotation effects, related to motion
in a non uniform field \cite{WRe} and a rotating AC field \cite{FClL},
respectively. Other applications include the control of agglomeration,
and the separation of proteins or living cells in suspension \cite{TLMa1}.
Nanorotors driven by a light force have also been studied \cite {MKh, TAs}.

Several methods have been used to obtain forces \cite{RTao,LGa,BJCo,Kang} and
torques \cite{FJGa,THi,JPHu,TLMa} in the past, some involving the use of an
interaction potential energy whose gradient is taken to obtain
the force \cite{RTao,KKi,FCl4,FCl5}. In this work we prove that if the particles
 are absorptive the
system is non-conservative and the net force experienced by each
member of the ensemble may not be derived from an interaction potential.
In fact, we show explicitly that structure in the interaction energy arising
from absorption resonances
in a pair of gold nanospheres exhibits energy minima leading to unphysical
equilibrium configurations that are not present if the force
is calculated directly from Coulomb's law \cite{RFu,FCl5}. Away from such resonances when absorption is negligible either method may be used leading to similar results.

In order to obtain explicit expressions for the force and torque we asume
the particles to be spherical, thus allowing a multipolar analysis and a
comparison with results obtained in the dipole approximation. For an AC
external excitation we find the dipole approximation to give accurate results if the
center to center separation between neighbors is not smaller than three
particle radii, while at closer interparticle distance the inclusion of
all multipoles gives rise to several resonances in the force and torque
strength, shifted to lower frequencies owing to particle-particle couplings.
This is in accordance with previous results on the electric excitation of
dielectric particles arrays showing a similar distance dependent
behavior \cite{FClA,FClT,RRo,JGe}.
Location of such resonances in the frequency spectrum  may be useful in
applications when the force or torque strength becomes important.
Within the same model we find that the appearance of a torque causing
particle rotation requires that the particle be dissipative.

The paper is organized as follows. In Sec. \ref{sec:inter} we present compact
expressions for the time-averaged force and torque acting over a particle in
an arbitrary array of nanoparticles in a uniform AC electric field.
The very structure of the resulting expressions
reveals the need for dissipation in order for a torque to arise. The cases
of linear and circular polarization are discussed. In order to assess the
relevance of higher multipoles in both forces and torque, in Sec. \ref{sec:special}
we apply our model to two gold nanospheres in an electric field parallel or
perpendicular to the interparticle axis. In Sec. \ref{sec:discus} we prove
that the presence of dissipation makes the system non conservative, and in
Sec. \ref{sec:conclu} we present our conclusions. Finally, two appendices are
added to provide details of the calculations.

\section{forces and torques on interacting particles in an AC field%
}

\label{sec:inter}

We consider a system of nanoparticles embedded
 in a non absorptive dielectric medium, excited by an external AC electric
field of angular frequency $\omega$. The particles are uncharged and
their material response to a local electric field may in general be characterized
by a complex response function $\epsilon(\omega)$. The external field induces a dipole moment
on each particle, which in turn excites multipoles on every other
member of the ensemble owing to the non uniformity of the electric field
it produces at each particle site. For simplicity we shall assume in what follows
that the particles are of spherical shape.

As known, for a dilute system with average center-to-center
separation of the order of three times the particle radii or more, the accuracy of
the dipole approximation is acceptable and the effect
of higher multipoles may be neglected \cite{FClA}. In such case and if
only two particles are present, the electric force between them may be
simply obtained by direct application of the discrete form of Coulomb´s law, as described
in  reference \cite{RFu}. When
separations less than three particle radii become involved however, the effect of
higher multipoles must be included \cite{FClT,RRo}. The general form of Coulomb's law to be used is then,

\begin{equation}
\left\langle \vec{F}_{i}\right\rangle =\frac{1}{2}\mathrm{Re}%
\int\rho_{i}^{*}\left(\vec{\mathit{r}}\right)       \vec{\mathit{E}} \left(\vec{\mathit{r}}%
\right)d^{3}\vec{r}\;,  \label{eq:deff1}
\end{equation}

\noindent where $\left\langle \vec{F}_{i}\right\rangle$ is the time-averaged force on particle $i$,
$\rho_{i}^{*}\left(\vec{\mathit{r}}\right)$ is its charge density and $\vec{\mathit{E}}\left(\vec{r}\right)$
is the local electric field due to the external sources and
other particles in the ensemble. A rather lengthy calculation then yields the force cartesian components (see Appendix A),

\begin{eqnarray}
\langle F_{ix}\rangle & = & \mathrm{Re}\sum_{l}C_{li}\mathrm{Re}T_{li}\:,
\label{eq:Fx444} \\
\langle F_{iy}\rangle & = & \mathrm{Re}\sum_{l} C_{li}\mathrm{Im}T_{li}\:,
\label{eq:Fy444} \\
\langle F_{iz}\rangle & = & \mathrm{Re}\sum_{l}C_{li}\sum_{m=-l}^{l}\sqrt{(l-m)(l+m)%
}q_{lmi}q_{l-1,m,i}^{*}\;,  \label{eq:Fz444}
\end{eqnarray}

\noindent where the pole order index $l$ here and in what follows covers the range of integers ${1,\infty}$. In the above expressions the coefficient

\begin{equation}
C_{li}=\frac{2\pi}{\sqrt{(2l+1)(2l-1)}\alpha_{li}}  \label{eq:cn1}
\end{equation}

\noindent weights the strength with which the multipole of order $l$ contributes, with $\alpha_{li}$
the corresponding particle polarizability, a complex quantity if absorption is present. Also

\begin{equation}
q_{lmi}=\int\rho_{i}\left(\vec{r}\right)r^{l}Y_{lm}^{*}\left(\theta,\phi%
\right)d^{3}{\vec{r}}\;  \label{eq:defq1}
\end{equation}

\noindent is the induced multipole of indices $l,m$ on particle $i$. and

\begin{equation}
T_{li}=\sum_{m}\sqrt{(l-m)(l-m-1)}q_{lmi}q_{l-1,m+1,i}^{\ast }\;.
\label{eq:cn2}
\end{equation}

\noindent $Y_{lm}$ is the usual complex spherical harmonic function. Methods to obtain the multipoles $q_{lmi}$ for arbitrary
configurations are described in Refs. \cite{FClT} and \cite{RRo}. Notice that since the force involves products of multipoles of different order, if there is
a single spherical particle and the external field is uniform only the dipole moment is excited and the force is zero.

Spinning of coupled particles in an external field has been observed in the past \cite{TLMa,GJSim}. In order
to capture this effect we consider next the time-averaged torque on sphere i
due to the local field, as given by

\begin{equation}
\left\langle \vec{\tau}_{i}\right\rangle =\frac{1}{2}\mathrm{%
Re\int\rho_{i}^{*}\left(\vec{\mathit{r}}\right)\vec{r}\times\vec{\mathit{E}}}%
\left(\vec{r}\right)d^{3}\vec{r}\;.  \label{eq:deftor}
\end{equation}

\noindent where the origin is taken at the particle center. Work similar to that done above for the forces (see Appendix B)
leads to the time-averaged torque cartesian components

\begin{eqnarray}
\langle\tau_{ix}\rangle & = & \mathrm{Im}\sum_{l=1}^{\infty}D_{li}\mathrm{Re}S_{li}\;,
\label{eq:taux1} \\
\langle\tau_{iy}\rangle & = & \mathrm{Im}\sum_{l=1}^{\infty}D_{li}\mathrm{Im}S_{li}\;,
\label{eq:tauy1} \\
\langle\tau_{iz}\rangle & = & \mathrm{Im}\sum_{l=1}^{\infty}D_{li}\sum_{m=-l}^{l}m%
\left|q_{lmi}\right|^{2}\:,  \label{eq:tauz1}
\end{eqnarray}

\noindent where the coefficients

\begin{equation}
D_{li}=\frac{2\pi}{(2l+1)\alpha_{li}}  \label{eq:cn3}
\end{equation}

\noindent are complex if $\alpha_{li}$ is, and

\begin{equation}
S_{li}=\sum_{m=-l}^{l-1}\sqrt{(l-m)(l+m+1)}q_{lmi}q_{l,m+1,i}^{*}\;.
\label{eq:defs}
\end{equation}

\noindent It is clear from Eqs. (\ref{eq:taux1}) to (\ref{eq:defs}) that if
the system has no dissipation, i.e. if $\alpha _{li}$ is real, the torque is
zero. We conclude that in general a torque arises in such systems from
dissipative electromagnetic interactions.

Even if there is dissipation
however, the torque may be suppressed by special symmetries. Such is the
case for a linear array subject to a uniform electric field parallel to the
line joining their centers. By choosing the $z$-axis to be aligned with this
line, only modes with $m=0$ are excited leading to zero torque, as may be
easily verified from the structure of the above equations. A similar
situation occurs if the applied electric field lies on the $xy$ plane since
in this case only modes with $m=\pm 1$ are excited symmetrically and the
torque is again zero. Nevertheless, it is worth noting that if the linear
array is under a uniform electric field with components along the $z$-axis
and the $xy$ plane, modes with $m=\pm 1$ and $m=0$ become excited. So,
according to Eqs. (\ref{eq:taux1}) to (\ref{eq:tauz1}) a torque is produced
provided that electromagnetic dissipation is not negligible. A similar
situation has been analyzed in Ref. \cite{TLMa} in the dipolar approach.

A torque does arise in such arrays also if they are subject to a rotating
electric field on the $xy$ plane. The field may be written as $\vec{E}%
=E_{0}(\pm \hat{x}-i\hat{y})e^{i\omega t}$ and the corresponding
coefficients of expansion of the potential are either $V_{1,+1}=\sqrt{2\pi /3%
}E_{0}(1-i)$ or $V_{1,-1}=\sqrt{2\pi /3}E_{0}(-1-i)$ depending of the sense
of rotation of the electric field vector given by the sign of the $x$
component \cite{Jac}. Correspondingly the excited modes are either $m=1$ or $%
m=-1$ and from Eq. (\ref{eq:tauz1}) it follows that a torque may appear. In
fact, from Eqs. (\ref{eq:alfalmi1}) and (\ref{eq:tauz1}) it can be shown
that for this case the time-average of the $z-$component of the torque is
given by

\begin{equation}
\langle\tau_{iz}\rangle=\frac{2\pi m}{a^{2l+1}}\sum_{l}\frac{l\;\mathrm{Im}%
\;\epsilon}{\left[l\;(\mathrm{Re}\;\epsilon-1)\right]^{2}+\left[l\;\mathrm{Im%
}\;\epsilon\right]^{2}}\left|q_{lmi}\right|^{2}\;.  \label{eq:tauz2}
\end{equation}

\noindent For the special case of a single sphere in a rotating external field
the torque is finite, in agreement with Refs.%
\cite{PRoD} and \cite{PLMa}.
The physical origin of such a torque is conservation of angular momentum.
The rotating field carries angular momentum, which is transferred to the
particles when absorption takes place causing them to experience a spinning
torque. Also, as noted in Ref. \cite{TLMa} when  a linearly polarized field is not
aligned with a symmetry axis of a linear array such as a pair, the local field
at each particle site has a rotating component, and the same argument applies.

\section{special case: two particles}

\label{sec:special}

We shall apply our general results to the simplest case, that of two
identical spheres of radii $a$ subject to a uniform oscillating
electric field, both parallel and perpendicular to a line joining the
spheres centers, that we choose to be the $z$ axis. These conditions will be
referred to as parallel and perpendicular excitation, respectively. In
computing the force we found convenient to use Eq. (\ref{eq:Fz3}) in Appendix A with the
replacement $V_{lmi}=b_{lmi}$, since the uniform external field produces no
direct force. Using relation (\ref{eq:blmi1}) then leads to,

\begin{equation}
\langle F_{iz}\rangle=-\frac{1}{2}\mathrm{Re}\sum_{lm}\sum_{l^{\prime}m^{%
\prime}}\sum_{j\neq i}(-1)^{l^{\prime}}A_{lmi}^{l^{\prime}m^{\prime}j}\sqrt{%
\frac{(2l+1)}{(2l-1)}(l-m)(l+m)}q_{l^{\prime}m^{\prime}j}q_{l-1,m,i}^{*}\;,
\label{eq:Fz5}
\end{equation}

\noindent where the coefficient $A_{lmi}^{l^{\prime}m^{\prime}j}$ that
couples multipoles in different particles is given by Eq. (\ref{eq:Almi}) in Appendix A.

\subsection{Parallel excitation}

In this geometry $\vec{E}=E_{0}e^{i\omega t}\hat{z}$ and modes with $m=0$
become excited only, yielding a force along the z-axis. From Eqs. (\ref{eq:Fx444}), (\ref{eq:Fy444}) and (\ref%
{eq:cn2}) it is seen that the time-averaged value of the components $x$ and $%
y$ of the force is zero, as expected from symmetry considerations. Using
Eq. (\ref{eq:Fz5}) we find after some algebra the force component
on sphere 1 centered at the origin

\begin{equation}
\langle F_{1z}\rangle=-2\pi\mathrm{Re}\sum_{ll^{\prime}}\left(-1\right)^{l^{%
\prime}}\frac{\left(l+l^{\prime}+1\right)!}{l!l^{\prime}!\sqrt{%
\left(2l+1\right)\left(2l^{\prime}+1\right)}R^{l+l^{\prime}+2}}%
q_{l,0,1}^{*}q_{l^{\prime},0,2}\;,  \label{eq:Fz6}
\end{equation}
where the multipole moments may be obtained using the formalism of Ref.\cite%
{RRo}. Here $R$ is the center to center distance between the two spheres.
The dipole approximation applies keeping the first term in this series,$%
(l=l^{\prime}=1)$, and the result agrees with that in Ref.\cite{RFu} as it should.

\subsection{Perpendicular excitation}

In this case the external field is in the $xy$ plane, and the external
potential in Eq. (\ref{eq:v4}) of Appendix A is expressed as $V^{ext}=V_{1,1,i}rY_{1,1}%
\left(\theta,\phi\right)+V_{1,-1,i}rY_{1,-1}\left(\theta,\phi\right)$ with $%
V_{1,\pm1}=\sqrt{2\pi/3}\left(\pm E_{x}-iE_{y}\right)$. The coupling
coefficients in Eq. (\ref{eq:Almi}) are null unless $m=m^{\prime}=\pm1$.
From Eq. (\ref{eq:Fz5}) we get this time,

\begin{equation}
\langle F_{1z}\rangle=2\pi\mathrm{Re}\sum_{ll^{\prime}}\left(-1\right)^{l^{%
\prime}}\frac{\left(l+l^{\prime}+1\right)!}{l!l^{\prime}!R^{l+l^{\prime}+2}}%
\sqrt{\frac{ll^{\prime}}{\left(2l+1\right)\left(2l^{\prime}+1\right)(l+1)%
\left(l^{\prime}+1\right)}}\left[q_{l,1,1}^{*}q_{l^{%
\prime},1,2}+q_{l,-1,1}^{*}q_{l^{\prime},-1,2}\right]\;.  \label{eq:Fz7}
\end{equation}

\noindent Keeping just the $l=l^{\prime}=1$ term in the series the dipole
approximation is obtained, which agrees with the corresponding expression in Ref.%
\cite{RFu}.

\subsection{Numerical Results}

We next show some numerical results for our test case of two particles. We
use a Drude dielectric function with parameters $\epsilon_{b}=9.9$, $%
\hbar\omega_{p}=8.2$ eV, $\Gamma=0.053$ eV, appropriate for gold
nanospheres \cite{CNo}. Figure 1 shows the average force for parallel
(solid line) as well as perpendicular (dashed line) excitation. One particle
is at the origin, while the other is at $z=R$. The separation is $R=2.005a$
and we have included multipoles up to order $L=40$ in the computation,
following the convergence criterion given in Ref. \cite{FClT}. The force
acting on the particle at the origin is attractive (positive) in the
parallel configuration and repulsive (negative) in the perpendicular
geometry, as expected. Three multipolar resonances are clearly resolved at
this separation, with force peaks greatly enhanced, about three orders of
magnitude above the background value. As the separation between the
particles is increased the resonances move to higher frequencies, decrease
in size and fewer of them become resolved \cite{RRo}. At a center to center
separation of about three particle radii and larger, only one resonance is
seen. This dipolar peak, at separation $R=3a$ and parallel excitation, has
been included in the figure for comparison with an amplification factor of one thousand (dash-dotted curve).

In Figure 2 we show the $z$ component of the average torque acting on each
nanoparticle as given by Eq. (\ref{eq:tauz2}). Separations are $R=2.005a$
(solid curve) and $R=3a$ (dashed curve). The pair is subject to an electric
field whose direction rotates in the plane $xy$. As for the force, several
resonances are resolved at small separation, while beyond about separation $%
R=3a$ only one peak is observed. It can be seen that as the spheres become
closer other resonances occur at frequencies below the single sphere dipole
resonance value $\omega=\omega_{p}/\sqrt{\epsilon_{b}+2}$. These additional
resonance frequencies correspond to resonant modes associated with the
multipole moments $q_{lmi}$.

\section{Limits in the use of an interaction energy to obtain the force}

\label{sec:discus}

The existence of dissipation makes a system non conservative. To see this,
recall that for ideal electromagnetic arrays where dissipation is absent,
the force acting on a particle may be obtained as the gradient with respect
to the particle coordinates of the configuration energy $W$. If the particle
makes a virtual displacement $\delta\xi$ the corresponding electric force it
is subject to is $F_{e}=\partial W_{e}/\partial\xi$, an expression obtained
by the energy balance equation,

\begin{equation}
\delta W_{source}=F_{e}\delta\xi+\delta W_{e}\;,  \label{eq:dW1}
\end{equation}

\noindent where $\delta W_{source}$ is the energy supplied by the sources to
maintain the potentials of the electrodes fixed, and $\delta W_{e}$ is the
variation in the energy stored in the field. It can be shown that for this
case $\delta W_{source}=2\delta W_{e}$ so that the expression $%
F_{e}=\partial W_{e}/\partial\xi$ is obtained \cite{Jac,WGr}. Nevertheless,
for real systems dissipation effects must be taken into account and that is
done adding a term $\delta W_{loss}$ in the right side of Eq. (\ref{eq:dW1}%
). This term depends on the path followed during the virtual displacement
since the polarization in the particle does and the energy loss is
determined by its imaginary part. If the particle is brought from point $A$
to point $B$, to the mechanical work done one must add the energy loss term $%
\int_{0}^{\tau}\overline{P}_{abs}dt$, where $\overline{P}_{abs}$ is the time
averaged power absorbed by the system and $\tau$ the time taken during the
displacement. Both the integrand and the upper limit of this integral depend
on the path making the mechanical system non conservative.

Based on the above argument we state that in a dissipative system it is
incorrect to obtain the force as the gradient of a potential. To illustrate
the difference between a direct application of Coulomb's law and the use of
a potential we consider two polarizable spheres of radius $a$, a distance $R$
apart in an electric field of frequency $\omega$ and amplitude $E_{0}$ which
for simplicity we choose to be parallel to the line joining the centers. In
the dipole approximation the interaction energy is of the form \cite{FCl5}

\begin{equation}
W_{int}(R)=U_{0}-\frac{1}{2}\mathrm{Re}[\beta_{1}(R)-\beta)]a^{3}E_{0}^{2}\;,
\label{eq:uint}
\end{equation}

\noindent where $U_{0}$ is the free-field interaction energy, $%
\beta=(\epsilon-1)/(\epsilon+2)$ with $\epsilon$ being the frequency
dependent dielectric function of the spheres, $\beta_{1}(R)=\beta/(1-\beta/4%
\sigma^{3})$, and $\sigma=R/2a$. Differentiating the second term in Eq. (\ref%
{eq:uint}) to get the force induced by the external field we obtain

\begin{equation}
F_{w}(R)=-\frac{a^{2}E_{0}^{2}}{48\sigma^{4}} \mathrm{Re} \frac{1}{(n-u)^{2}}
\;,  \label{eq:felew}
\end{equation}

\noindent where $n=(1-1/4\sigma^{3})/3$ and the complex spectral variable $%
u=1/(\epsilon-1)$ has been used. By contrast, if the direct Coulomb's method
is used one gets \cite{RFu}

\begin{equation}
F_{c}(R)=-\frac{a^{2}E_{0}^{2}}{48\sigma^{4}} \frac{1}{|n-u|^{2}} \;.
\label{eq:felec}
\end{equation}

\noindent The two forms (\ref{eq:felew}) and (\ref{eq:felec}) agree only
when the dielectric function is real, and dissipation is absent. In Figure 3
we compare the force obtained using these two expressions for a pair of gold
nanospheres with a dielectric function as described en Sec. III. As can be
observed while the direct Coulomb's method gives an attractive force at all
frequencies, the model based on the gradient of the interaction energy
presents two peaks and an unphysical change of sign in the force.

\section{conclusions}

\label{sec:conclu}

We have shown that in an ensemble of polarizable spheres in an oscillating
electric field, the presence of a rotation torque requires the particle
material to be dissipative. We also show that energy loss due to dissipation
makes the system non conservative so that it is improper to use an
interaction energy to derive the force, an approach that has been employed
erroneously in the past \cite{KKi}. Our results are an extension of previous
work done for the case of an isolated pair using the dipolar model \cite{RFu}. When
interparticle distances are shorter than three particle radii it is known
that the dipole approximation is not adequate, and higher multipoles must be
considered \cite{FClA,RRo}. Electromagnetic resonances associated with such
multipoles are known to appear, that should have a mirror spectrum in the
forces and torques as well. We have explicitely shown this to be the case in the simple case of a
pair.

\begin{acknowledgments}
During the elaboration of this paper one of the contributing authors,
Professor Ronald Fuchs of Ames Laboratory and Iowa State University, has
passed away. This work is dedicated to him. One of us (PR) thanks to
Escuela de Ingenier\'{\i}a El\'ectrica, Pontificia Universidad Cat\'olica de Valpara%
\'{\i}so for its support.
\end{acknowledgments}

\newpage{}

\appendix
%dummy comment inserted by tex2lyx to ensure that this paragraph is not empty
%dummy comment inserted by tex2lyx to ensure that this paragraph is not empty

\section{Time-averaged force}

We consider en ensemble of N spheres in the presence of an external electric field.
Choosing a coordinate system with origin at the center of
particle $i$, the electric potential at a point in the medium due to the polarizd spheres is given by
\cite{Jac}

\begin{equation}
V\left(\vec{r}\right)=\sum_{l=1}^{\infty}\sum_{m=-l}^{+l}\frac{4\pi}{2l+1}%
q_{lmi}\frac{Y_{lm}\left(\theta,\phi\right)}{r^{l+1}}+\sum_{l=1}^{\infty}%
\sum_{m=-l}^{+l}\sum_{j=1}^{N}\frac{4\pi}{2l+1}q_{lmj}\frac{Y_{lm}\left(\bar{%
\theta_{j}},\bar{\phi_{j}}\right)}{\bar{R_{j}}^{l+1}}\;,  \label{eq:v1}
\end{equation}

\noindent where the multipole moment of order $l,m$ in particle j has been defined in Eq. (\ref{eq:defq1}). The center of sphere $j$ is
at $\vec{R}_{j}$ and $\vec{r}-\vec{R}_{j}=\left(\bar{R}_{j},\bar{\theta}_{j},%
\bar{\phi}_{j}\right)$ is the position vector of the observation point with
respect to the center of sphere $j$. To uncouple vectors $\vec{r}$ and $\vec{%
R}_{j}$ we use the identities \cite{FClL2},

\begin{equation}
\frac{Y_{lm}(\bar{\theta}_{j},\bar{\phi}_{j})}{\bar{R}_{j}^{l+1}}=(-1)^{l+m}%
\left[\frac{2l+1}{4\pi(l+m)!(l-m)!}\right]^{1/2}\left[\frac{\partial}{%
\partial x}+i\frac{\partial}{\partial y}\right]^{m}\frac{\partial^{l-m}}{%
\partial z^{l-m}}\frac{1}{\left|\vec{r}-\vec{R}_{j}\right|}\;,
\label{eq:id1}
\end{equation}

\begin{equation}
\frac{1}{\left|\vec{r}-\vec{R}_{j}\right|}=\sum_{l=0}^{\infty}\frac{r_{<}^{l}%
}{r_{>}^{l+1}}\frac{4\pi}{2l+1}\sum_{m=-l}^{+l}(-1)^{m}Y_{lm}(\theta,%
\phi)Y_{l,-m}(\theta_{j},\phi_{j})\;,  \label{eq:id2}
\end{equation}

\begin{equation}
\frac{\partial^{n}}{\partial z^{n}}Y_{lm}(\theta,\phi)r^{l}=\left[\frac{2l+1%
}{(2l-2n+1)}\frac{(l+m)!}{(l+m-n)!}\frac{(l-m)!}{(l-m-n)!}\right]%
Y_{l-n,m}(\theta,\phi)r^{l-n}\;,  \label{eq:id3}
\end{equation}

\begin{equation}
\left[\frac{\partial}{\partial x}+i\frac{\partial}{\partial y}\right]%
^{p}Y_{lm}(\theta,\phi)r^{l}=\left[\frac{2l+1}{(2l-2p+1)}\frac{(l-m)!}{%
(l-m-2p)!}\right]Y_{l-p,m+p}(\theta,\phi)r^{l-p}\;.  \label{eq:id4}
\end{equation}

\noindent In Eq. (\ref{eq:id2}) $r_{<}(r_{>})$ is the lower (higher) value
between $r=\left|\vec{r}\right|$ and $R_{j}=\left|\vec{R}_{j}\right|$; Eq. (%
\ref{eq:id3}) is valid for $l\geq n$ and $\left|m\right|\leq l-n$ while Eq. (%
\ref{eq:id4}) is valid for $l\geq p$ and $-l\leq m\leq l-2p$. From Eqs. (\ref%
{eq:id1}) to (\ref{eq:id4}) and adding the potential $V^{ext}$ due to the
external field, Eq. (\ref{eq:v1}) becomes

\begin{equation}
V\left(\vec{r}\right)=\sum_{l,m}\frac{4\pi}{2l+1}q_{lmi}\frac{%
Y_{lm}\left(\theta,\phi\right)}{r^{l+1}}+\sum_{l,m}b_{lmi}Y_{lm}(\theta,%
\phi)r^{l}+V^{ext}\;,  \label{eq:v2}
\end{equation}

\noindent where
\begin{equation}
b_{lmi}=\sum_{l^{\prime}m^{\prime}}\sum_{j\neq
i}A_{lmi}^{l^{\prime}m^{\prime}j}q_{l^{\prime}m^{\prime}j}\;.
\label{eq:blmi1}
\end{equation}

\noindent Here $A_{lmi}^{l^{\prime}m^{\prime}j}$ is the coupling coefficient
between $q_{lmi}$ and $q_{l^{\prime}m^{\prime}j}$ (with $i\neq j$) \cite{RRo}

\begin{eqnarray}
A_{lmi}^{l^{\prime}m^{\prime}j} & = & \left(-1\right)^{m^{\prime}}\frac{%
Y_{l+l^{\prime},m-m^{\prime}}^{*}\left(\theta_{ij},\phi_{ij}\right)}{%
\left|R_{ij}\right|^{l+l^{\prime}+1}}  \nonumber \\
& & \times\left[\frac{\left(4\pi\right)^{3}\left(l+l^{\prime}+m-m^{\prime}%
\right)!\left(l+l^{\prime}-m+m^{\prime}\right)!}{\left(2l+1\right)\left(2l^{%
\prime}+1\right)\left(2l+2l^{\prime}+1\right)\left(l+m\right)!\left(l-m%
\right)!\left(l^{\prime}+m^{\prime}\right)!\left(l^{\prime}-m^{\prime}%
\right)!}\right]^{1/2},  \label{eq:Almi}
\end{eqnarray}

\noindent and $\vec{R}_{i}-\vec{R}_{j}=\left(R_{ij},\theta_{ij},\phi_{ij}%
\right)$. Equations (\ref{eq:v2}) to (\ref{eq:Almi}) are general and valid
for any array of spherical particles and arbitrary direction of the applied
electric field. All expressions here and below are given in Gaussian units.

In order to obtain the average force we use Eq. (\ref{eq:deff1}) making the replacement $E\left(\vec{\mathit{r}}\right)=-\nabla V_{i}(\vec{r})$ for
the local electric field due to the polarized system. Here
\begin{equation}
V_{i}\left(\vec{r}\right)=\sum_{lm}b_{lmi}r^{l}Y_{lm}\left(\theta,\phi%
\right)+V^{ext}\left(\vec{r}\right)\;.
\end{equation}

\noindent If we expand the external potential as
\begin{equation}
V^{ext}\left(\vec{r}\right)=\sum_{lm}V_{lmi}^{ext}r^{l}Y_{lm}\left(\theta,%
\phi\right)\;,  \label{eq:v4}
\end{equation}

\noindent the above equation may be written in the form
\begin{equation}
V_{i}\left(\vec{r}\right)=\sum_{lm}V_{lmi}r^{l}Y_{lm}\left(\theta,\phi%
\right)\;,  \label{eq:v3}
\end{equation}

\noindent where $V_{lmi}=V_{lmi}^{ext}+b_{lmi}$.

In order to obtain explicit expressions for the components of the force we
first write the spherical harmonics in the above equation in terms of
Legendre functions using the relation
\begin{equation}
Y_{lm}\left(\theta,\phi\right)=\sqrt{\frac{(2l+1)}{4\pi}\frac{%
\left(l-m\right)!}{\left(l+m\right)!}}P_{l}^{m}(\cos\theta)e^{im\phi}\;.
\label{eq:Ylm1}
\end{equation}
Then Eq. (\ref{eq:v3}) may be recast as
\begin{equation}
V_{i}\left(\vec{r}\right)=\sum_{lm}D_{lmi}r^{l}P_{l}^{m}\left(\cos\theta%
\right)e^{im\phi}\;,  \label{eq:v5}
\end{equation}
where
\begin{equation}
D_{lmi}=V_{lmi}\sqrt{\frac{(2l+1)}{4\pi}\frac{\left(l-m\right)!}{%
\left(l+m\right)!}}\;.  \label{eq:defd1}
\end{equation}
Therefore the spherical components of the electric field are
\begin{eqnarray}
E_{r} & = & -\frac{\partial V_{i}(\vec{r})}{\partial r}=-%
\sum_{lm}D_{lmi}lr^{l-1}P_{l}^{m}(\xi)e^{im\phi},  \label{eq:Er1} \\
E_{\theta} & = & -\frac{1}{r}\frac{\partial V_{i}(\vec{r})}{\partial\theta}%
=-\sum_{lm}D_{lmi}r^{l-1}\frac{\partial}{\partial\theta}P_{l}^{m}(\xi)e^{im%
\phi}=\sum_{lm}D_{lmi}r^{l-1}\sqrt{1-\xi^{2}}\frac{\partial}{\partial\xi}%
P_{l}^{m}(\xi)e^{im\phi}  \label{eq:Eteta1} \\
E_{\phi} & = & -\frac{1}{r\sin\theta}\frac{\partial V_{i}(\vec{r})}{%
\partial\phi}=-\sum_{lm}D_{lmi}r^{l-1}\frac{1}{\sqrt{1-\xi^{2}}}%
P_{l}^{m}(\xi)ime^{im\phi}.  \label{eq:Efi1}
\end{eqnarray}
In Eqs. (\ref{eq:Er1}) to (\ref{eq:Efi1}) we have defined $\xi=\mathrm{cos}%
\theta$. The corresponding Cartesian components of the electric field are
given by
\begin{eqnarray}
E_{x} & = &
E_{r}\sin\theta\cos\phi+E_{\theta}\cos\theta\sin\phi-E_{\phi}\sin\phi\;,
\label{eq:Ex1} \\
E_{y} & = &
E_{r}\sin\theta\sin\phi+E_{\theta}\cos\theta\cos\phi+E_{\phi}\cos\phi\;,
\label{eq:Ey1} \\
E_{z} & = & E_{r}\cos\theta-E_{\theta}\sin\theta\;.  \label{eq:Ez1}
\end{eqnarray}
It is useful to calculate linear combinations of $E_{x}$ and $E_{y}$ defined
as
\begin{eqnarray}
E_{+} & = & E_{x}+iE_{y}\;,  \label{eq:Emas1} \\
E_{-} & = & E_{x}-iE_{y}\;.  \label{eq:Emenos1}
\end{eqnarray}
Introducing relations (\ref{eq:Er1}) to (\ref{eq:Ey1}) into Eq. (\ref%
{eq:Emas1}) one obtains
\begin{equation}
E_{+}=-\sum_{lm}D_{lmi}r^{l-1}\left[l\sqrt{1-\xi^{2}}P_{l}^{m}(\xi)-\xi\sqrt{%
1-\xi^{2}}\frac{\partial}{\partial\xi}P_{l}^{m}(\xi)-\frac{m}{\sqrt{1-\xi^{2}%
}}P_{l}^{m}(\xi)\right]e^{i(m+1)\phi}\;.  \label{eq:Emas2}
\end{equation}
The relations
\begin{equation}
(1-\xi^{2})\frac{\partial P_{l}^{m}(\xi)}{\partial\xi}=(l+m)P_{l-1}^{m}(%
\xi)-l\xi P_{l}^{m}(\xi)\;,  \label{eq:id5}
\end{equation}
\begin{equation}
(l-m)P_{l}^{m}(\xi)-\xi(l+m)P_{l-1}^{m}(\xi)=\sqrt{1-\xi^{2}}%
P_{l-1}^{m+1}(\xi)\;,  \label{eq:id6}
\end{equation}
lead then to
\begin{equation}
E_{+}=-\sum_{lm}D_{lmi}r^{l-1}P_{l-1}^{m+1}(\xi)e^{i(m+1)\phi}\;.
\label{eq:Emas3}
\end{equation}
Using Eq. (\ref{eq:defd1}) and (\ref{eq:Ylm1}) one obtains
\begin{equation}
E_{+}=-\sum_{lm}V_{lmi}r^{l-1}\sqrt{\frac{2l+1}{2l-1}(l-m)(l-m-1)}%
Y_{l-1,m+1}(\theta,\phi)\;.  \label{eq:Emas4}
\end{equation}
Similarly, from Eqs. (\ref{eq:Emenos1}) and (\ref{eq:Er1}) to (\ref{eq:Ey1})
follows
\begin{equation}
E_{-}=-\sum_{lm}D_{lmi}r^{l-1}\left[l\sqrt{1-\xi^{2}}P_{l}^{m}(\xi)-\xi\sqrt{%
1-\xi^{2}}\frac{\partial}{\partial\xi}P_{l}^{m}(\xi)+\frac{m}{\sqrt{1-\xi^{2}%
}}P_{l}^{m}(\xi)\right]e^{i(m+1)\phi}\;.  \label{eq:Emenos2}
\end{equation}

\noindent The recurrence relation $\xi
P_{l-1}^{m}(\xi)-P_{l}^{m}(\xi)=(l+m-1)\sqrt{1-\xi^{2}}P_{l-1}^{m-1}(\xi)$
and Eq. (\ref{eq:id5}) can be used to find that

\begin{equation}
E_{-}=\sum_{lm}V_{lmi}r^{l-1}\sqrt{\frac{2l+1}{2l-1}(l-m)(l+m-1)}%
Y_{l-1,m-1}(\theta,\phi)\;.  \label{eq:Emenos3}
\end{equation}

\noindent To obtain $E_{z}$ we use the relation $(1-\xi^{2})\frac{\partial
P_{l}^{m}(\xi)}{\partial\xi}=(l+m)P_{l-1}^{m}(\xi)-l\xi P_{l}^{m}(\xi)$ ,
and Eqs. (\ref{eq:Er1}), (\ref{eq:Eteta1}) and (\ref{eq:Ez1}) to give

\begin{equation}
E_{z}=-\sum_{lm}V_{lmi}r^{l-1}\sqrt{\frac{2l+1}{2l-1}(l-m)(l+m)}%
Y_{l-1,m}(\theta,\phi)\;.  \label{eq:Ez2}
\end{equation}

\noindent The Cartesian components of the time-averaged force acting upon
sphere $i$ are then given by

\begin{eqnarray}
\langle F_{ix}\rangle & = & \frac{1}{2}\mathrm{Re}\int\rho_{i}^{*}(\vec{r}%
)E_{x}d^{3}\vec{r}=\frac{1}{2}Re\int\rho_{i}^{*}(\vec{r})\frac{1}{2}%
(E_{+}+E_{-})d^{3}\vec{r}  \nonumber \\
& = & -\frac{1}{4}\mathrm{Re}\int\rho_{i}^{*}(\vec{r})\sum_{lm}V_{lmi}r^{l-1}%
\sqrt{\frac{2l+1}{2l-1}}  \nonumber \\
& & \times\left[\sqrt{(l-m)(l-m-1)}Y_{l-1,m+1}(\theta,\phi)-\sqrt{%
(l+m)(l+m-1)}Y_{l-1,m-1}(\theta,\phi)\right]d^{3}\vec{r}  \nonumber \\
& = & -\frac{1}{4}\mathrm{Re}\sum_{lm}V_{lmi}\sqrt{\frac{2l+1}{2l-1}}
\nonumber \\
& & \times\left[\sqrt{(l-m)(l-m-1)}q_{l-1,m+1,i}^{*}-\sqrt{(l+m)(l+m-1)}%
q_{l-1,m-1,i}^{*}\right]\;,  \label{eq:Fx3}
\end{eqnarray}

\begin{eqnarray}
\langle F_{iy}\rangle & = & \frac{1}{2}\mathrm{Re}\int\rho_{i}^{*}(\vec{r}%
)E_{y}d^{3}\vec{r}=\frac{1}{2}Re\int\rho_{i}^{*}(\vec{r})\frac{i}{2}%
(-E_{+}+E_{-})d^{3}\vec{r}  \nonumber \\
& = & \frac{1}{2}\mathrm{Re}\int\rho_{i}^{*}(\vec{r})\frac{i}{2}%
\sum_{lm}V_{lmi}r^{l-1}\sqrt{\frac{2l+1}{2l-1}}  \nonumber \\
& & \times\left[\sqrt{(l-m)(l-m-1)}Y_{l-1,m+1}(\theta,\phi)+\sqrt{%
(l+m)(l+m-1)}Y_{l-1,m-1}(\theta,\phi)\right]d^{3}\vec{r}  \nonumber \\
& = & \frac{1}{4}\mathrm{Re}\; i\sum_{lm}V_{lmi}\sqrt{\frac{2l+1}{2l-1}}
\nonumber \\
& & \times\left[\sqrt{(l-m)(l-m-1)}q_{l-1,m+1,i}^{*}+\sqrt{(l+m)(l+m-1)}%
q_{l-1,m-1,i}^{*}\right]\;,  \label{eq:Fy3}
\end{eqnarray}

\begin{eqnarray}
\langle F_{iz}\rangle & = & \frac{1}{2}\mathrm{Re}\int\rho_{i}^{*}(\vec{r}%
)E_{z}d^{3}\vec{r}  \nonumber \\
& = & -\frac{1}{2}\mathrm{Re}\int\rho_{i}^{*}(\vec{r})\sum_{lm}V_{lmi}r^{l-1}%
\sqrt{\frac{2l+1}{2l-1}(l-m)(l+m)}Y_{l-1,m}(\theta,\phi)d^{3}\vec{r}
\nonumber \\
& = & -\frac{1}{2}\mathrm{Re}\sum_{lm}V_{lmi}\sqrt{\frac{2l+1}{2l-1}%
(l-m)(l+m)}q_{l-1,m,i}^{*}\;,  \label{eq:Fz3}
\end{eqnarray}

\noindent The coefficients $V_{lmi}$ and $q_{lmi}$ are related by \cite{FClT}

\begin{equation}
q_{lmi}=-\frac{2l+1}{4\pi }\alpha _{li}V_{lmi}\;,  \label{eq:qlmi2}
\end{equation}

\noindent where $\alpha _{li}$ is the multipole polarizability of the sphere
$i$ given by \cite{RRo}

\begin{equation}
\alpha _{li}=\frac{l(\epsilon -1)}{l(\epsilon +1)+1}a_{i}^{2l+1}\;.
\label{eq:alfalmi1}
\end{equation}

\noindent We next use relation (\ref{eq:qlmi2}) in Eqs. (\ref{eq:Fx3}), (\ref%
{eq:Fy3}) and (\ref{eq:Fz3}) to get the force components as a sum, bilinear
in the induced multipole moments. Using the property $q_{l,-m}^{*}=\left(-1%
\right)^{m}q_{lm}$ that arises from definition (\ref{eq:defq1}) and the
properties of spherical harmonics, one then gets

\begin{eqnarray}
\langle F_{ix}\rangle & = & \mathrm{Re}\sum_{l}C_{li}\mathrm{Re}T_{li}\:,
\label{eq:Fx445} \\
\langle F_{iy}\rangle & = & \mathrm{Re}\sum C_{li}\mathrm{Im}T_{li}\:,
\label{eq:Fy445} \\
\langle F_{iz}\rangle & = & \mathrm{Re}\sum_{l}C_{li}\sum_{m}\sqrt{(l-m)(l+m)%
}q_{lmi}q_{l-1,m,i}^{*}\;,  \label{eq:Fz445}
\end{eqnarray}

\noindent where
\begin{equation}
C_{li}=\frac{2\pi}{\sqrt{(2l+1)(2l-1)}\alpha_{li}}  \label{eq:cn11}
\end{equation}

\noindent is in general a complex quantity involving the polarizability $%
\alpha_{li}$, and

\begin{equation}
T_{li}=\sum_{m}\sqrt{(l-m)(l-m-1)}q_{lmi}q_{l-1,m+1,i}^{\ast }\;.
\label{eq:cn22}
\end{equation}

\noindent The force components are thus given in compact form, convenient
for numerical computation.

\newpage

\section{Time-averaged torque}

In this Appendix we derive general expressions for the time-averaged
components of the torque acting upon particle $i$ in a set of $N$ polarizable
spherical nanoparticles of radii $a$ in the presence of a uniform AC electric field.
This is given by

\begin{equation}
\left\langle \vec{\tau}_{i}\right\rangle =\frac{1}{2}\mathrm{%
Re\int\rho_{i}^{*}\left(\vec{\mathit{r}}\right)\vec{r}\times\vec{\mathit{E}}}%
\left(\vec{r}\right)d^{3}\vec{r}\;.  \label{eq:deftor1}
\end{equation}

The time averaged torque over particle i in the ensemble is given by Eq. (\ref{eq:deftor1}). The corresponding Cartesian components are

\begin{eqnarray}
\langle\tau_{ix}\rangle & = & \frac{1}{2}\text{Re}\int\rho_{i}^{*}(\vec{r}%
)(yE_{z}-zE_{y})d^{3}\vec{r}\;,  \label{eq:tauxA1} \\
\langle\tau_{iy}\rangle & = & \frac{1}{2}\text{Re}\int\rho_{i}^{*}(\vec{r}%
)(zE_{x}-zE_{z})d^{3}\vec{r}\;,  \label{eq:tauyA1} \\
\langle\tau_{iz}\rangle & = & \frac{1}{2}\text{Re}\int\rho_{i}^{*}(\vec{r}%
)(xE_{y}-yE_{x})d^{3}\vec{r}\;.  \label{eq:tauzA1}
\end{eqnarray}

\noindent Using the field and distance variables defined as $%
E_{\pm}=E_{x}\pm iE_{y}$ and $r_{\pm}=x\pm iy\;$, the $x$ and $y$ components
of the torque can be expressed as

\begin{eqnarray}
\langle\tau_{ix}\rangle & = & \frac{1}{4}\text{Re}\int\rho_{i}^{*}(\vec{r}%
)i(W_{+}-W_{-})d^{3}\vec{r}\;,  \label{eq:tauxA2} \\
\langle\tau_{iy}\rangle & = & \frac{1}{4}\text{Re}\int\rho_{i}^{*}(\vec{r}%
)(W_{+}+W_{-})d^{3}\vec{r}\;,  \label{eq:tauyA2}
\end{eqnarray}

\noindent where $W_{+}=zE_{+}-r_{+}E_{z}$ and $W_{-}=zE_{-}-r_{-}E_{z}$.
From Eqs. (\ref{eq:Emas3}) and (\ref{eq:Ez2}) for $E_{+}$and $E_{z}$
respectively, and introducing relation (\ref{eq:Ylm1}) we have

\begin{eqnarray}
zE_{+} & = & -r\xi\sum_{lm}D_{lmi}r^{l-1}P_{l-1}^{m+1}(\xi)e^{i(m+1)\phi}
\label{eq:zEmasA1} \\
r_{+}E_{z} & = & -r\sqrt{1-\xi^{2}}e^{i\phi}%
\sum_{lm}D_{lmi}r^{l-1}P_{l-1}^{m}(\xi)e^{i(m)\phi}\;.  \label{eq:rEzA1}
\end{eqnarray}

\noindent Eqs. (\ref{eq:zEmasA1}), (\ref{eq:rEzA1}) and the identity $-\xi
P_{l-1}^{m+1}(\xi)+(l+m)\sqrt{1-\xi^{2}}P_{l-1}^{m}(\xi)=-P_{l}^{m+1}(\xi)$
lead to

\begin{eqnarray}
W_{+} & = & \sum_{lm}D_{lmi}r^{l}\left[-\xi P_{l-1}^{m+1}(\xi)+(l+m)\sqrt{%
1-\xi^{2}}P_{l-1}^{m}(\xi)\right]e^{i(m+1)\phi}  \nonumber \\
& = & \sum_{lm}V_{lmi}\sqrt{\frac{(2l+1)}{4\pi}\frac{(l-m)!}{(l+m)!}}r^{l}%
\left[-P_{l}^{m+1}(\xi)\right]e^{i(m+1)\phi}  \nonumber \\
& = & -\sum_{lm}V_{lmi}r^{l}\sqrt{(l-m)(l+m+1)}Y_{l,m+1}(\theta,\phi)\;.
\label{eq:wmasA3}
\end{eqnarray}

\noindent Using Eqs. (\ref{eq:Emenos3}) and (\ref{eq:Ez2}) for $E_{-}$and $%
E_{z}$ respectively, and introducing relation (\ref{eq:Ylm1}) we have
\begin{eqnarray}
zE_{-} & = &
r\xi\sum_{lm}D_{lmi}r^{l-1}(l+m)(l+m-1)P_{l-1}^{m-1}(\xi)e^{i(m-1)\phi}\;,
\label{eq:zEmenosA1} \\
r_{-}E_{z} & = & -r\sqrt{1-{\xi}^{2}}e^{-i\phi}%
\sum_{lm}D_{lmi}r^{l-1}(l+m)P_{l-1}^{m}(\xi)e^{im\phi}\;.  \label{eq:rEzA2}
\end{eqnarray}

\noindent Eqs. (\ref{eq:zEmenosA1}) , (\ref{eq:rEzA2}) and the identity $%
\xi(l+m-1)P_{l-1}^{m-1}(\xi)+\sqrt{1-\xi^{2}}P_{l-1}^{m+1}(%
\xi)=(l-m+1)P_{l}^{m-1}(\xi)$ leads to

\begin{eqnarray}
W_{-} & = & \sum_{lm}D_{lmi}r^{l}(l+m)\left[\xi(l+m-1)P_{l-1}^{m-1}(\xi)+%
\sqrt{1-\xi^{2}}P_{l-1}^{m}(\xi)\right]e^{i(m-1)\phi}\;,  \nonumber \\
& = & \sum_{lm}V_{lmi}\sqrt{\frac{(2l+1)}{4\pi}\frac{(l-m)!}{(l+m)!}}\;
r^{l}(l+m)(l-m+1)P_{l}^{m-1}(\xi)e^{i(m-1)\phi}\;,  \nonumber \\
& = & \sum_{lm}V_{lmi}r^{l}\sqrt{(l+m)(l-m+1)}Y_{l,m-1}(\theta,\phi)\;,
\label{eq:wmenosA3}
\end{eqnarray}

\noindent Using Eqs. (\ref{eq:wmasA3}), (\ref{eq:wmenosA3}) and the
definition $q_{lmi}=\int\rho_{i}\left(\vec{r}\right)r^{l}Y_{lm}^{*}\left(%
\theta,\phi\right)d^{3}{\vec{r}}\;$ we get

\begin{eqnarray}
\left\langle \tau_{ix}\right\rangle & = & \frac{1}{4}\text{Re}%
\int\rho_{i}^{*}(\vec{r})i(W_{+}-W_{-})d^{3}\vec{r}\;,  \nonumber \\
& = & -\frac{1}{4}\mathrm{Re}\sum_{lm}iV_{lmi}\left[\sqrt{(l-m)(l+m+1)}%
q_{l,m+1,i}^{*}+\sqrt{(l+m)(l-m+1)}q_{l,m-1,i}^{*}\right]\;.
\label{eq:tauxA4}
\end{eqnarray}

\noindent From the relation $q_{lmi}=-\frac{2l+1}{4\pi}\alpha_{lmi}V_{lmi}$
between the multipole moment $lm$ induced in particle $i$ and the
corresponding expansion coefficient $V_{lmi}$ we obtain

\begin{equation}
\langle\tau_{ix}\rangle=\mathrm{Re}\sum_{l}\frac{i\pi}{(2l+1)\alpha_{li}}%
\left[S_{li}+S_{li}^{*}\right]\;,  \label{eq:tauxA5}
\end{equation}

\noindent where

\begin{equation}
S_{li}=\sum_{m=-l}^{l-1}\sqrt{(l-m)(l+m+1)}q_{lmi}q_{l,m+1,i}^{*}\;.
\label{eq:defsA1}
\end{equation}

\noindent A similar development for the $y$ component of the torque gives
\begin{equation}
\langle\tau_{iy}\rangle=\mathrm{Re}\sum_{l}\frac{\pi}{(2l+1)\alpha_{li}}%
\left[S_{li}-S_{li}^{*}\right]\;.  \label{eq:tauyA3}
\end{equation}

\noindent The $z$ component of the torque may be rewritten as

\begin{equation}
\langle\tau_{iz}\rangle=\frac{1}{4i}\mathrm{Re}\int\rho_{i}^{*}(\vec{r})%
\left[r_{-}E_{+}-r_{+}E_{-}\right]d^{3}\vec{r}\;.  \label{eq:tauzA2}
\end{equation}

\noindent Using definitions of $r_{+}$ and $r_{-}$ and relations (\ref%
{eq:Emas4}) and (\ref{eq:Emenos3}) for $E_{+}$ and $E_{-}$ we get

\begin{equation}
r_{-}E_{+}-r_{+}E_{-}=-\sum_{lm}D_{lmi}r^{l}\left[\sqrt{1-\xi^{2}}%
P_{l-1}^{m+1}(\xi)+(l+m)(l+m-1)\sqrt{1-\xi^{2}}P_{l-1}^{m-1}(\xi)\right]%
e^{im\phi}.  \label{eq:difA1}
\end{equation}

\noindent Using $(l+m-1)\sqrt{1-\xi^{2}}P_{l-1}^{m-1}(\xi)=\xi
P_{l-1}^{m}(\xi)-P_{l}^{m}(\xi)$, the expression between square brackets in
Eq. (\ref{eq:difA1}) , which we denote by $C$ becomes
\begin{eqnarray}
C & = & \sqrt{1-\xi^{2}}P_{l-1}^{m+1}(\xi)+(l+m)\left[\xi
P_{l-1}^{m}(\xi)-P_{l}^{m}(\xi)\right]\;,  \nonumber \\
& = & (l+m)\xi P_{l-1}^{m}(\xi)+\sqrt{1-\xi^{2}}P_{l-1}^{m+1}(%
\xi)-(l+m)P_{l}^{m}(\xi)\;.  \label{eq:defcA2}
\end{eqnarray}

\noindent Since $(l+m)\xi P_{l-1}^{m}(\xi)+\sqrt{1-\xi^{2}}%
P_{l-1}^{m+1}(\xi)=(l-m)P_{l}^{m}(\xi)$ we get
\begin{eqnarray}
C & = & (l-m)P_{l}^{m}(\xi)-(l+m)P_{l}^{m}(\xi)\;,  \nonumber \\
& = & -2mP_{l}^{m}(\xi)\;.  \label{eq:defcA4}
\end{eqnarray}

\noindent Using Eqs. (\ref{eq:tauzA2}), (\ref{eq:difA1}) and (\ref{eq:defcA4}%
) we find
\begin{eqnarray}
\left\langle \tau_{iz}\right\rangle & = & \frac{1}{2}\text{Re}%
\int\rho_{i}^{*}(\vec{r})\frac{1}{2i}(r_{-}E_{+}-r_{+}E_{-})d^{3}\vec{r}\;,
\nonumber \\
& = & \frac{1}{2}\mathrm{Re}\int\rho_{i}^{*}(\vec{r})\frac{1}{2i}%
\sum_{lm}V_{lmi}r^{l}2mY_{lm}(\theta,\phi)d^{3}\vec{r}\;.  \label{eq:tauzA4}
\end{eqnarray}

\noindent With the definition $q_{lmi}=\int\rho_{i}\left(\vec{r}%
\right)r^{l}Y_{lm}^{*}\left(\theta,\phi\right)d^{3}{\vec{r}}\;$ and the
relation $q_{lmi}=-\frac{2l+1}{4\pi}\alpha_{lmi}V_{lmi}$ for eliminating $%
V_{lmi}$ we obtain our final result for the $z$ component of the torque
\begin{eqnarray}
\left\langle \tau_{iz}\right\rangle & = & \frac{1}{2}\text{Re}\frac{1}{i}%
\sum_{lm}V_{lmi}mq_{lmi}^{*}\;,  \nonumber \\
& = & \mathrm{Re}\sum_{lm}\frac{2\pi i}{2l+1}\frac{m}{\alpha_{li}}%
q_{lmi}q_{lmi}^{*}\;.  \label{eq:tauzA6}
\end{eqnarray}

\newpage{}

\newpage{}

\section*{Figures}

\begin{figure}[h]
\centering{}\includegraphics{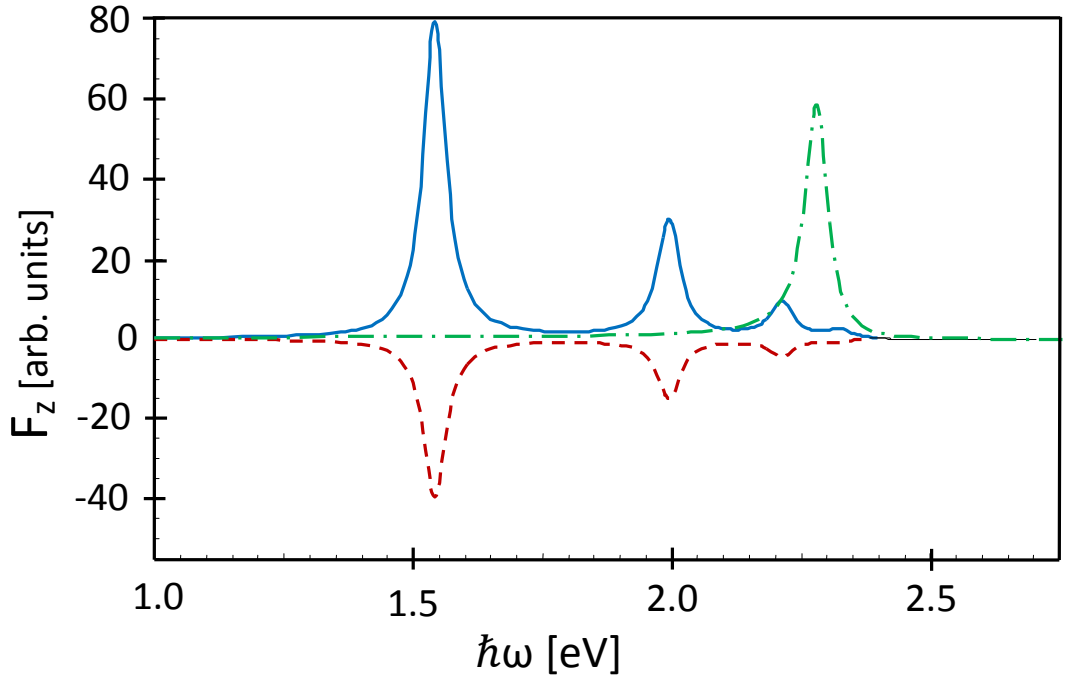}
\caption{Electric force between two identical gold nanospheres as a function
of frequency of the applied field, with separation $2.005a$ between their
centers. The solid (dashed) curve corresponds to parallel (perpendicular)
excitation calculated including multipoles up to $L=40$. The dash-dotted
curve corresponds to the average force calculated for parallel excitation
and separation $3a$, with an amplification factor $1000$.}
\label{fig:two}
\end{figure}

\begin{figure}[h]
\centering{}\includegraphics{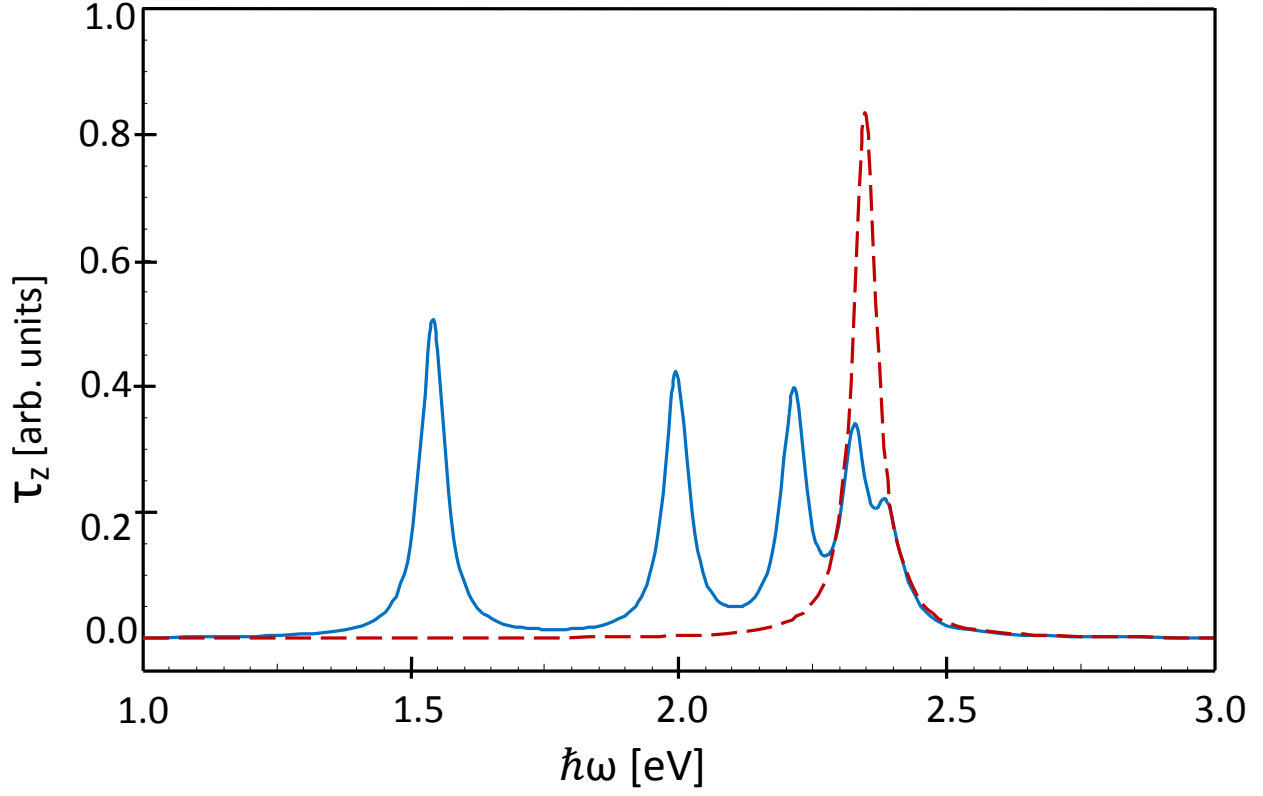}
\caption{Time averaged torque acting on a particle for a system of two gold
nanospheres subjected to a rotating electric field,
as a function of frequency. Separation between their centers
are $2.005a$ (solid curve) and $3a$ (dashed curve). Results were obtained
including multipoles up to $L=40$ and $L=10$, respectively.}
\label{fig:three}
\end{figure}

\begin{figure}[h]
\centering{}\includegraphics{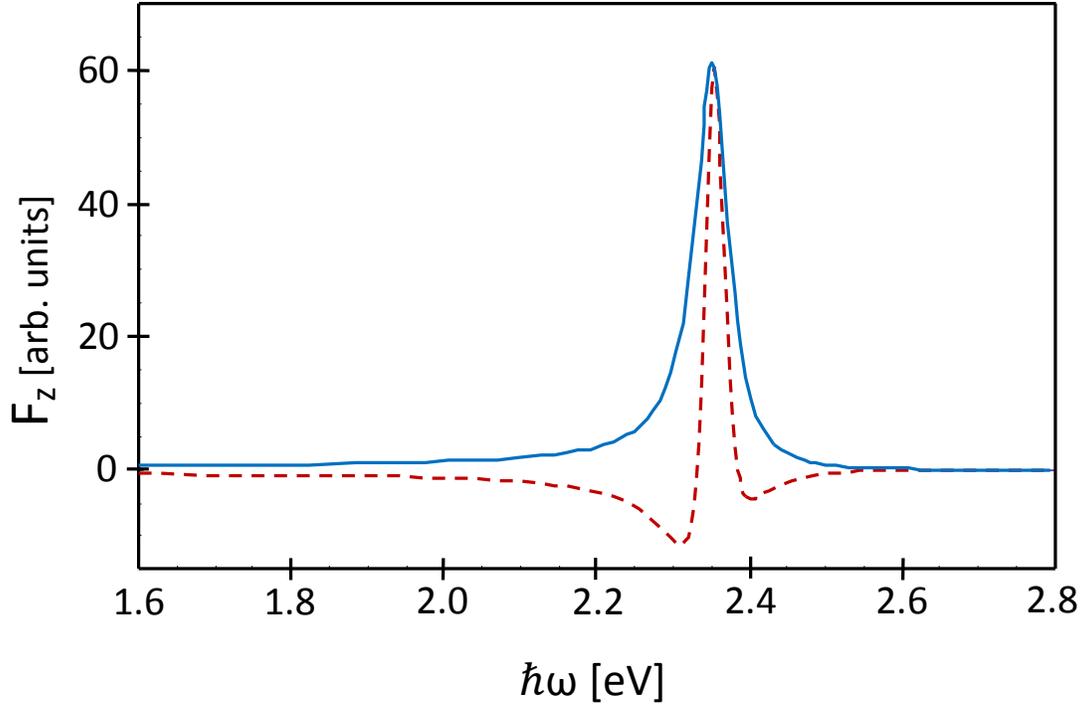}
\caption{Force between two identical gold nanospheres in the parallel
configuration as a function of frequency, with separation $3a$ between their
centers. Solid and dashed curves correspond to the force calculated from
Coulomb's law and using the derivative of an interaction potential,
respectively.}
\label{fig:four}
\end{figure}

\end{document}